# Experimental demonstration of a thermal-EM concentrator for enhancing EM signals and converging heat fluxes simultaneously

Hanchuan Chen,[1, a)] Yichao Liu,[1, a)] Fei Sun,[1, b)] Qianhan Sun[1)] Xiaoxiao Wu,[2)] and Ran Sun[1)]

[1] *Key Lab of Advanced Transducers and Intelligent Control System, Ministry of Education and Shanxi Province, College of Electronic Information and Optical Engineering, Taiyuan University of Technology, Taiyuan 030024, China*

[2] *Advanced Materials Thrust, The Hong Kong University of Science and Technology (Guangzhou), Guangzhou 510000, China*

[a] *Hanchuan Chen and Yichao Liu contributed equally to this work.*

[b] *Corresponding author E-mail addresses: sunfei@tyut.edu.cn*

**Simultaneously concentrating EM waves and heat fluxes to the same target region within an on-chip system carries substantial academic research importance and practical application value. Nevertheless, existing researches are primarily aimed at the design and experimentation of concentrators for individual EM waves or temperature fields. In this work, a thermal-EM concentrator, capable of simultaneously concentrating EM waves and heat fluxes, is designed using transformation optics/thermodynamics and fabricated with engineered EM-thermal metamaterials. The concentrating effects of the proposed thermal-EM concentrator on the thermal fluxes and EM waves are verified through numerical simulations and experimental measurements, respectively, which are in good agreement with each other. Both numerically simulated and experimentally measured results demonstrate the concentrating capability of the proposed thermal-EM concentrator, which can concentrate broadband TM-polarized EM waves ranging from 8-12 GHz and heat/cold flows to the same target region within an on-chip operating environment. The thermal-EM concentrator exhibits a thermal focusing efficiency close to 100% and more than three times enhancement of the magnetic field at the designed center frequency of 10 GHz. The proposed thermal-EM concentrator can be utilized for efficient cooling for the specified component and simultaneously enhancing the EM antenna's radiation/reception efficiency within an on-chip system.**

Enhancing electromagnetic (EM) field, especially broadband collecting of EM energy, has significant applications in various fields, such as improving the efficiency of wireless power transfer,[1–3] detecting/harnessing extremely weak EM signals,[4–7] producing non-linear effects,[8] and so on. Meanwhile, concentrating hot/cold fluxes to a specified region also has important applications in thermal management, such as collecting the heat converted from solar cells,[9] and efficiently cooling a given area on a chip. In many scenarios, there is a requirement to simultaneously collect/concentrate broadband EM waves and heat/cold flows in the same target region. As an example, in on-chip intelligent systems, there is a need to efficiently focus broadband EM signals on the on-chip antenna while simultaneously concentrating the cold flow generated by the on-chip semiconductor cooler to the antenna's location. This prevents the antenna from generating heat by absorbing EM signals, which could lead to localized overheating of the on-chip system. However, there is still a lack of studies on thermal-EM concentrators (TEC) that can simultaneously achieve effective concentrating of EM waves on the surface of a chip while concentrating the heat/cold flow within the chip.

In previous studies, several different structures that can individually focus/concentrate EM waves or temperature fields have been proposed, such as, EM concentrators using transformation optics,[10–13] thermal concentrators designed by transformation thermodynamics,[9,14,15] EM lenses and compression structures by zero refractive index materials,[16–22] and Janus thermal device with thermal focusing capability by thermal metamaterials.[23–28] With the advancement of metamaterials, structures capable of effective focusing in both electrostatic and temperature fields have been proposed and experimentally validated.[29–33] However, unlike the electrostatic field, which satisfies the Laplace equation similar to the temperature field, the propagation of EM waves follows the Helmholtz equation. Therefore, the realization of a TEC for both EM waves and the temperature field simultaneously remains challenging, and still lacks theoretical design and experimental verification.

In recent years, multi-physics metamaterials that are effective for two or more physical fields have been emerging. Meanwhile, structures for simultaneously controlling multiple physical fields are continuously appearing, such as electrostatic-thermal cloaks,[31–33] EM-acoustic cloaks,[34] magnetostatic-acoustic stealth coats[35] and EM-acoustic lenses.[36] In our recent research, we explore the properties of thermal-EM null medium (TENM), which can simultaneously project the distributions of both electromagnetic waves and temperature fields along its main axis, and propose a standardized black-box designing method known as thermal-EM surface transformation, enabling the simultaneous control of temperature fields and electromagnetic waves. This work provides theoretical foundation and design approach for the simultaneous dual-physics field control of EM waves and temperature fields[37]. In this study, a TEC, which can simultaneously concentrate TM-polarized EM waves and heat fluxes into the same target region, is designed and fabricated by using engineered EM-thermal metamaterials. Both simulated and measured results show that the proposed TEC can enhance electromagnetic signals in the target region over a relatively wide frequency band (i.e., 8-12GHz) around the designed operating central frequency ($f_0$=10 GHz) and concentrate the heat fluxes (e.g., hot/cold flow) into the same target region.

Fig. 1 is an example of the application of the proposed TEC in an on-chip system. As shown in Fig. 1(a), for an integrated circuit chip, the on-chip antenna located at the center serves to receive/radiate the EM signal and generate heat during its operation. To prevent the heat generated by the on-chip antenna from affecting surrounding integrated components (e.g., the adjacent 6G communication module), semiconductor cooling elements are commonly incorporated in the on-chip system (indicated by the blue rectangular structure at the top and bottom of the diagram, with the cold flow represented by blue arrows). In such a working environment, if the chip has a high integration density, two inevitable issues will arise. Firstly, due to the high integration density, only a small amount of cold flow produced by semiconductor cooling elements can reach the cooling target (e.g., the 6G communication module in Fig. 1(a)), while a significant portion disperses to other surrounding modules. Therefore, it is unable to achieve efficient cooling for the specified component. Secondly, with the increase in chip integration density, the effective area for on-chip antennas to radiate/receive EM signals is becoming smaller. Therefore, there is an urgent need for an EM focusing/concentrating lens that can effectively concentrate on-chip EM signals in the region where the on-chip antenna is located, to enhance the on-chip antenna's EM radiation/reception efficiency. Hence, a TEC capable of simultaneously concentrating EM waves on the chip and the heat/cold flows within the chip can effectively address both issues in a highly integrated on-chip system. As depicted in Fig. 1(b), when the TEC is introduced around the target region where the communication module with an antenna is located, it can simultaneously concentrate the cold flow and EM signals to its target region, thereby achieving efficient cooling for the specified component and enhancing the EM antenna's radiation/reception efficiency simultaneously. Movie 1 demonstrates the functionality of the proposed TEC in a high-integration on-chip system.

Here, we show the theoretical design method of the proposed TEC. From the perspective of coordinate transformation, considering two cylindrical regions with a common radius of $c$ in the reference space (Fig. 2(a)) and the real space (Fig. 2(b)), the external transformation of the two cylindrical regions is kept as an identity transformation. This ensures that the medium outside the cylindrical regions ($r' > c$) in both spaces are the same, assumed to be a uniform background medium (i.e., $\kappa = \kappa_0$, $\varepsilon = \varepsilon_0$, $\mu = \mu_0$). Additionally, the TEC is obtained by performing a continuous coordinate transformation inside two cylindrical regions of the same radius $c$. Specifically, a purple annular region in the reference space (Fig. 2(a)) is stretched to a green annular region in the real space (Fig. 2(b)). At the same time, the core with a radius of $a$ inside the purple annular region is compressed into a smaller core with a radius of $b$ in the real space. Since the core in the real space is compressed from a much larger core in the reference space, this results in the EM waves and temperature fields inside the core in the real space also being compressed, thereby achieving concentrating effects on both EM waves and temperature fields. In particular, as the area of the purple annular region in the reference space approaches zero, the green annular region in the real space becomes the TENM, which can guide both temperature field and EM wave in radial direction to the central core. A segmented continuous linear transformation is used for each region in the above coordinate transformation, and with the help of simple calculations in transformation optics[38–41] and transformation thermodynamics,[33,42,43] the material parameters of the corresponding TEC in Fig. 2(b) can be obtained as: $\kappa = \kappa_0$, $\varepsilon = \varepsilon_0$, $\mu = \mu_0$ for $r' > c$; $\kappa_\rho = \varepsilon_\rho \to \infty$, $\kappa_\varphi = \varepsilon_\varphi = \mu_z \sim 0$. for $b < r' < c$; $\kappa_\rho = \kappa_\varphi = \kappa_0$, $\varepsilon_\rho = \varepsilon_\varphi = \varepsilon_0$, $\mu_z = \mu_0(c/b)^2$ for $r' < b$, considering only the required materials for TM-polarized EM waves.

As shown in Fig. 2(c), the engineered EM-thermal metamaterials are designed with effective parameters as close as possible to the theoretical requirements for thermal conductivity and electromagnetic parameters, aiming to implement the proposed TEC. For the background medium in the region $r' > c$, 13W commercial thermal conductive pad with a thermal conductivity $\kappa = \kappa_0 = 13$ W/(mK) is used in our following experiment to perform as an on-chip background environment in Fig. 1. As the thermal conductive pad acts as dielectric (i.e., $\varepsilon = 6.5\varepsilon_0$, $\mu = \mu_0$) for EM waves at the designed operating frequency ($f_0 = 10$ GHz, $\lambda_0 = 30$ mm), and it's the thickness is approximately $0.06\lambda_0$ (on the deep subwavelength order), its impact on EM waves is approximately equivalent to that of air. For the green annular region $b < r' < c$, the TENM can be realized by staggered wedge-shaped copper and expanded polystyrene (EPS) arrays with $\theta_1 = \theta_2 = 5°$ (i.e., the same filling factors for copper and EPS) and a height of $H = 0.17\lambda_0$ above the background thermal conductive pad, which can perform as a reduced TENM with effective material parameters of $\kappa_\rho = 146.7$ W/(mK), $\varepsilon_\rho \to \infty$, $\kappa_\varphi = 3.7$ W/(mK), $\varepsilon_\varphi = 3.28\varepsilon_0$, $\mu_z = 0.64\mu_0$. The detailed derivation of the effective parameters for the reduced TENM using effective medium theory can be found in Supplementary Note 1. For the central core $r' < b$, considering that its material parameters need to simultaneously satisfy the thermal conductivity $\kappa_\rho = \kappa_\varphi = \kappa_0$ and electromagnetic parameters $\varepsilon_\rho = \varepsilon_\varphi = \varepsilon_0$, $\mu_z = \mu_0(c/b)^2$, high dielectric ceramic disc is used. Its thermal conductivity ($\kappa=0.65$W/(mK)) is close to that of air, while the electromagnetic parameters approximately satisfy $\varepsilon = 9.2\varepsilon_0 \approx \varepsilon_0(c/b)^2$, $\mu_z = \mu_0$, rather than the ideal EM parameters ($\varepsilon_\rho = \varepsilon_\varphi = \varepsilon_0$, $\mu_z = \mu_0(c/b)^2$) calculated from transformation optics. The reason for this simplification is that previous studies on EM concentrators have shown that simplifying the electromagnetic parameters in this manner can significantly reduce electromagnetic wave scattering under the wavelength satisfying FP resonant conditions. At the same time, it ensures that the central region of the concentrator still has a strong electromagnetic field enhancement effect, even though the magnetic field distribution at the core is no longer uniform.[11] More information regarding thermal conductivity and electromagnetic parameters of materials used in the experiment can be found in the Supplementary Note 2.

The photograph of the TEC sample used in the experiments are shown in Fig. 2(d). In the experiment, the thermal conductive pad is manually cut into a rectangular shape with an area of 200*250mm (colored blue in Fig. 2(d)) to simulate the on-chip working environment in Fig. 1. For the annular TENM region, 36 copper wedges (colored orange in Fig. 2(d)) are machined by wire cutting, and 36 EPS wedges (colored white in Fig. 2(d)) are manually made by thermal cutting. Subsequently, these copper and EPS wedges are manually assembled and glued together to form the required ring structure. For the central core, the high dielectric ceramic disc is precisely machined into the desired shape by computer numerical control machining, with material parameters of $\varepsilon = 9.2\varepsilon_0$, $\mu = \mu_0$, $\kappa = 0.65$ W/(mK). To measure the magnetic field within the core during the experiment, a subwavelength cylindrical air hole with a radius of $r_0 = 0.1\lambda_0$ and a height of $h = 0.1\lambda_0$ is carved into the middle of the central core for placing the EM probe. The whole sample is fixed on a 200*250*50 mm EPS block to hold the sample, which is then placed on a $10\lambda_0$ high EPS block to lift it off the ground and to prevent interference from the ground on the EM signals. For the designed operating central frequency ($f_0 = 10$ GHz) in the microwave band, EPS can perform as free space (i.e., static air) for EM waves. At the same time, the thermal conductivity of EPS ($\kappa = 0.04$W/(mK)) is very small, which can also be treated as static air for the process of heat conduction. More information about sample preparation can be found in Supplementary Note 3 and Movie 2.

To verify the simultaneous concentrating effect of the fabricated sample on EM waves and temperature fields, we perform experimental measurements of the temperature field and electromagnetic field, respectively, and their corresponding test environments are shown in Fig. 3(a) and (b). For temperature field measurements in Fig. 3(a), thermal conductive pad is placed around the TEC sample as the background material, with its two ends connected to heat sink (i.e., the cooling source) and positive temperature coefficient heater (PTCH) to establish directed cold flow incident on the sample. The whole system is situated within a vacuum chamber to prevent air convection. The temperature field distribution around the sample is observed/recorded using an infrared camera (IR Camera, FOTRIC 288). To minimize the impact of varying surface emissivity of different regions of the sample on the temperature field observation, a layer of Teflon tape with a stabilized surface emissivity of 0.95 is applied onto the sample's surface (see Fig. 2(d)). As part of a thermal comparison experiment, the TEC sample at the center position is removed, leaving only a complete thermal conductive pad, with all other configurations unchanged. More information about the preparation and measurement for the thermal experiment can be found in Supplementary Note 4.

The three-dimensional (3D) simulated and experimental measured results of the normalized steady-state temperature field distribution with/without the TEC are shown in Figs. 4(a,b) and Figs. 4(c,d), respectively. In the 3D thermal simulations, the dimensions of the simulated sample are the same as those of the fabricated one, and all environmental settings in simulations are similar to the experimental settings. In the simulations, the boundaries of the computational domain are thermally insulated (corresponding to the experimental condition where the sample is situated in a vacuum chamber). The left side of the thermal conductive pad is set at a fixed temperature of 293.15K (cooling source), while the right side is fixed at a temperature boundary of 393.15K, which corresponds to the heat sink and the PTCH in experiment, respectively. The 3D simulated and measured temperature distributions with the TEC in Figs. 4(a,c) are identical, indicating that the temperature field is effectively concentrated in the central core. In comparison, the 3D simulated and measured temperature distributions without the TEC in Figs. 4(b,d) also match very well. However, once the TEC is removed, there is no thermal concentration effect in the central core. Both simulated and measured results verify that when the cold flow incidents onto the TEC, it will also be effectively concentrated into the central core, thereby improving the cooling/dissipation efficiency of components in the central region. More information about temperature field simulation can be found in Supplementary Note 5.

To quantitatively examine the thermal concentrating performance of the fabricated TEC, the thermal concentrating efficiency (*CE*) is often introduced as[9]:

$$CE = \frac{|T|_{x=b} - T|_{x=-b}|}{|T|_{x=c} - T|_{x=-c}|}, \tag{1}$$

where $T_{x=\pm b}$ and $T_{x=\pm c}$ are the measured temperatures at locations $x=\pm b$ and $x=\pm c$, respectively. For the simulated and measured results in Fig. 4(a) and 4(c), the thermal *CE*s are both close to 100%. Specific *CE*s for different $b$ and $c$ can be found in the Supplementary Note 6.

For the EM field measurements in Fig. 3(b), the whole sample (including both TEC and background thermal conductive pad) is situated within a microwave chamber. The vector network analyzer (VNA, R&S ZVL13) is utilized for the generation and measurement of microwave EM signals. An open-loop probe (performing as a loop antenna with a diameter of $0.2\lambda_0$) is connected to Port 1 of the VNA, positioned $0.33\lambda_0$ away from the side edge of the sample to generate the TM-polarized EM wave. Additionally, another open-loop probe (also with a diameter of $0.2\lambda_0$) is attached to Port 2 of the VNA, serving as a detection probe and placed within the air hole at the center of the sample to measure the magnetic field. To avoid the influence of ground reflection on EM waves, the sample is placed on a $10\lambda_0$ high EPS block, which has a dielectric constant close to air. As part of an EM comparison experiment, the TEC sample is removed, leaving only a complete thermal conductive pad, with all other configurations unchanged. More information about the preparation and measurement for the EM experiment can be found in Supplementary Note 7.

The 3D simulated normalized intensity of magnetic field distribution with/without TEC are shown in Fig. 5(a) and (b), respectively. In the 3D EM simulations, the dimensions of the simulated sample are the same as those of the fabricated one, which are the same as these parameters used in thermal simulations. Scattering boundary condition is applied around the air domain to absorb EM waves and simulate an infinite free space. An open-loop antenna with a 6 mm diameter is placed on the left side of the TEC, and a surface current of 1A/m is applied along the loop to excite the incident TM-polarized EM wave. For the design operating central frequency of 10 GHz, compared to the case without using TEC in Fig. 5(b), when the designed TEC is applied in Fig. 5(a), the magnetic field in the central core is significantly enhanced.

More information about EM simulation can be found in Supplementary Note 5.

To quantitatively describe the degree of magnetic field enhancement, the enhancement factor $A$ is defined as:

$$A = \left|\frac{H_{TEC}}{H_0}\right|, \qquad (2)$$

where $H_{TEC}$ and $H_0$ represent the magnetic fields at the center of the core with and without the TEC, respectively. Fig. 5(c) shows the relationship between the simulated (red line) and experimentally measured (blue line) enhancement factors as a function of frequency ranging from 8-12 GHz. The curves of the measured enhancement factors closely coincide with the simulated curves at almost all frequency points, demonstrating excellent agreement (with minor differences mainly attributed to sample fabrication errors). The simulate enhancement factors are slightly elevated compared to the measured enhancement factors across the entire scanning frequency range. This discrepancy arises from the excitation of the incident magnetic field in the simulation, achieved by imposing a fixed current of 1A/m on the loop coil, leading to uncontrollable output power and an inability to maintain a constant output power as in the experimental setup. The measured results in Fig. 5(c) indicate that the fabricated TEC sample can achieve broadband magnetic field enhancement in the frequency range of 8-12 GHz. Moreover, at the designed center frequency of 10 GHz, more than three times enhancement can be obtained.

In summary, a TEC is designed using transformation optics/thermodynamics to simultaneously achieve concentrate EM waves on the surface of a chip and focus the heat/cold flow within the chip. Subsequently, engineered EM-thermal metamaterials are designed to realize the proposed TEC, which is then fabricated and experimentally measured. Both simulated and measured results demonstrate that the proposed TEC can concentrate TM-polarized EM waves and heat flow to the same target region. Moreover, the TEC exhibits a thermal focusing efficiency close to 100% and a broadband magnetic field enhancement effect for microwave signals ranging from 8-12 GHz. The proposed TEC is highly compatible with existing integrated circuit technologies and may find significant applications in next-generation on-chip systems that require broadband EM signal enhancements and targeted cooling simultaneously.

**Data availability**
The main data and models supporting the findings of this study are available within the paper and Supplementary Information. Further information is available from the corresponding authors upon reasonable request.

**Acknowledgments**
This work is supported by the National Natural Science Foundation of China (Nos. 12274317, 12374277, and 61971300), Basic Research Project of Shanxi Province 202303021211054, and University Outstanding Youth Foundation of Taiyuan University of Technology.

**Competing interests**

The authors declare no competing financial interests.

**Captions**

Figure 1 (Color online) Schematic diagram of an on-chip system. The transmission situations of cold flows and EM signals are shown in cases (a) without TEC and (b) with TEC, respectively. (a) only a small part of cold flows and EM signals can reach the target region (e.g., the central 6G communication module together integrated with EM antennas). In this scenario, achieving efficient cooling for the specified target region is not possible, and the strength of EM signals radiated from or received by antennas is very weak. (b) when the TEC is introduced around the target region where the communication module integrated with EM antennas is located, it can simultaneously concentrate the cold flows and EM signals to its target region. In this scenario, achieving efficient cooling for the specified target region and enhancing the EM antenna's radiation/reception efficiency can occur simultaneously.

Figure 2 (a) and (b) are the reference space and the real space in transformation optics/thermodynamics. (c) Schematic diagram of the TEC. TEC is divided into the core region (yellow region) and the shell region (red and blue annular band regions). Among them, the core region is made of high dielectric ceramic disc with radius $b$ and a cylindrical hole with radius $r_0$ and height $h$ hollowed out at the center. The outer radius of the shell region is $c$, of which the blue region is EPS, the central angle of wedge-shape EPS is $\theta_1$, the red region is copper, and the central angle of wedge-shape copper is $\theta_2$. The thickness of the TEC is $H$. (d) The photo of the experimentally fabricated TEC. The sample is constructed based on the structure outlined in (c) with the following structural parameters: $b$ = 30 mm, $r_0$ =3 mm, $h$ = 3 mm, $c$ = 90 mm, $\theta_1$ = $\theta_2$ = 5°, $H$ = 5 mm. During the experimental measurement, the sample is fixed on a 200*250*50 mm rectangular EPS board. In the thermal measurement experiment, the entire surface of the sample is covered with Teflon tape to ensure a constant emissivity of 0.95, facilitating the observation of temperature distribution by an IR camera. To reveal the internal structure of the sample, only half of the area is covered with Teflon tape in the photo (the black portion in the photo).

Figure 3 Diagram of the experimental testing environment. (a) Diagram of the experimental testing environment for the temperature field. In the temperature field experiment, the ambient temperature is maintained at room temperature 293.15 K via air conditioning. The right PTCH is energized by electric wires as a constant high temperature boundary. The left Heat sink provides a constant low temperature cooling source. Thermal conductive pad with thermal conductivity 13 W/(mK) are placed around the sample to simulate the on-chip environment. A layer of Teflon tape with a stabilized surface emissivity of 0.95 (see Fig. 2(d) is covered on the top surface of the sample to ensure the accuracy of the temperature distribution

data collected by the IR Camera. The whole system is situated within a vacuum chamber. (b) Diagram of the experimental testing environment for the EM field. In the EM field experiment, the sample is placed on a $10\lambda_0$ high EPS block. The VNA Ports 1 and 2 are connected to coaxial cables with open-loop probes (performing as a loop antenna with a diameter of $0.2\lambda_0$). Probe 1 is placed $\lambda_0$ away from the side of the sample to excite the TM-polarized electromagnetic wave, and Probe 2 is placed within the air hole at the center of the sample to measure the magnetic field by VNA.

Figure 4  Verification of the temperature field concentrating effect by the proposed TEC through 3D simulation and experimental measurement. (a) 3D simulation of temperature field distribution with TEC. (b) 3D simulation of temperature field distribution without TEC. (c) Experimental measurement of temperature field distribution with TEC. (d) Experimental measurement of temperature field distribution without TEC.

Figure 5  Verification of the EM field concentrating effect by the proposed TEC through 3D simulation and experimental measurement. (a) 3D simulation of magnetic field amplitude distribution with TEC. (b) 3D simulation of magnetic field amplitude distribution without TEC. (c) The enhancement factor *A* of the TEC is obtained from both simulation (red curve) and experimental measurement (blue curve) across the 8-12GHz frequency range.

Figure 1

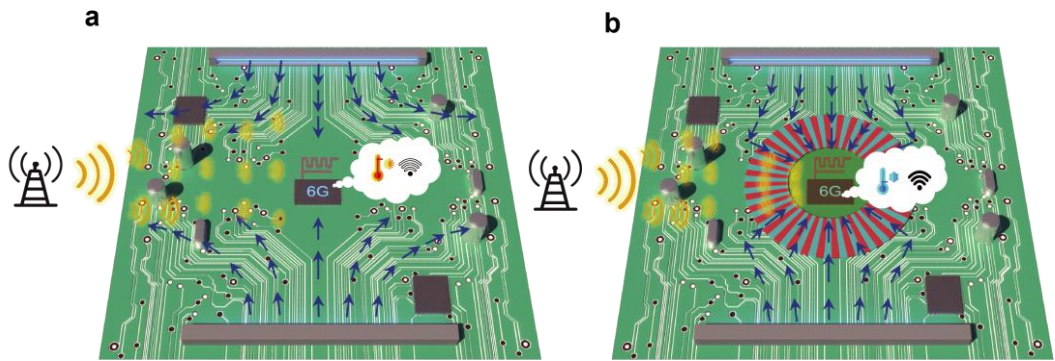

Figure 2

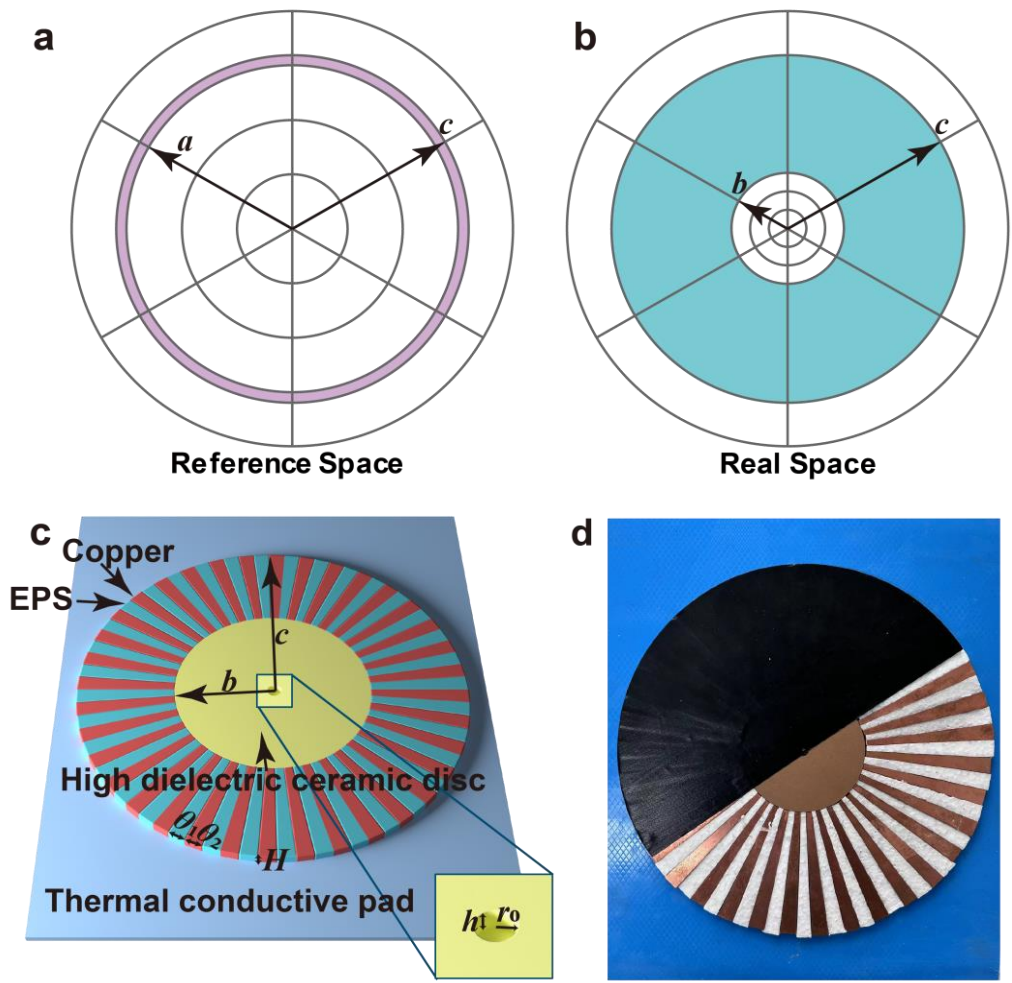

Figure 3

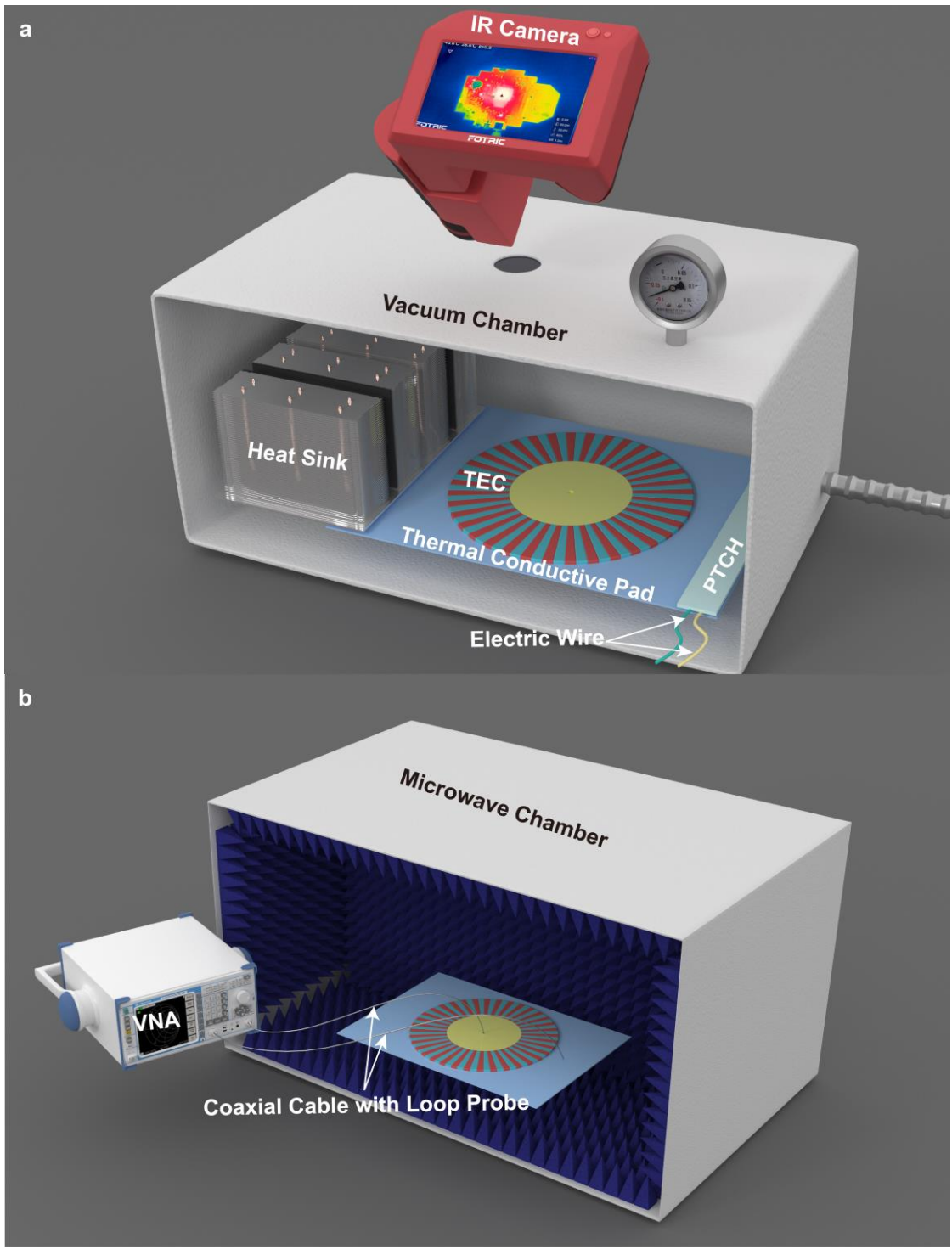

Figure 4

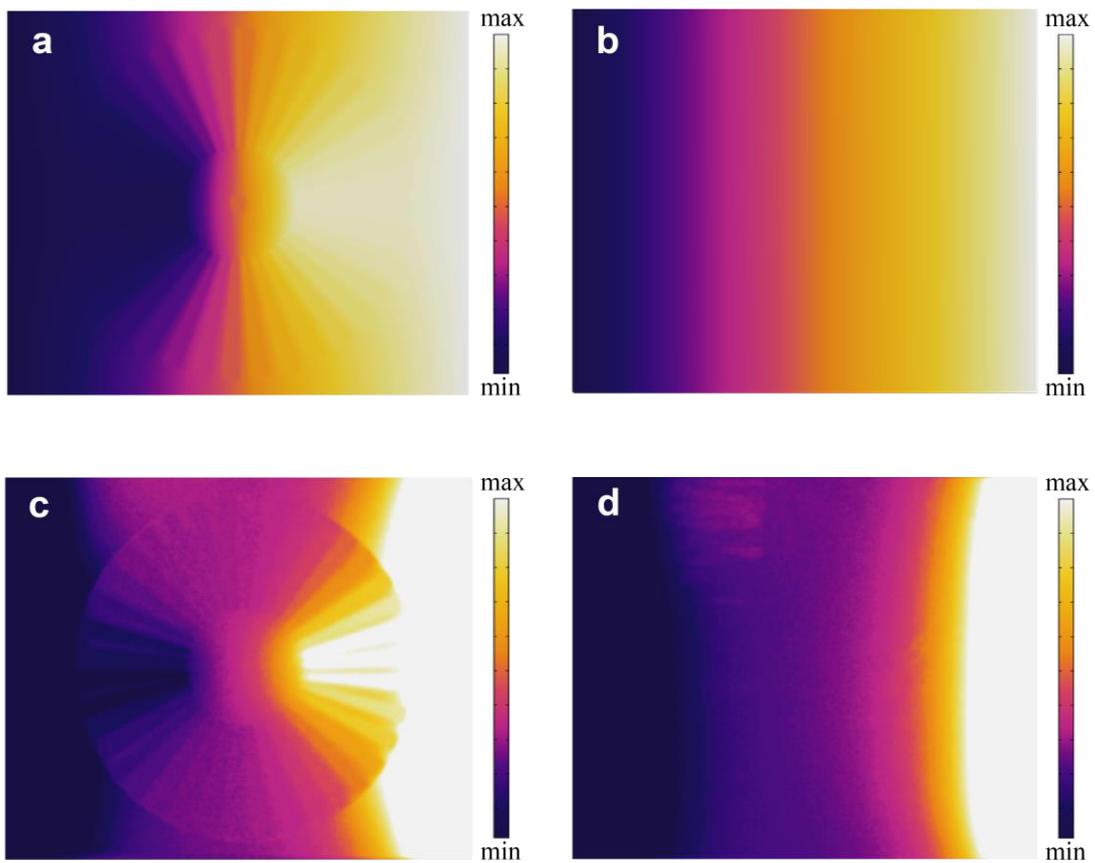

Figure 5

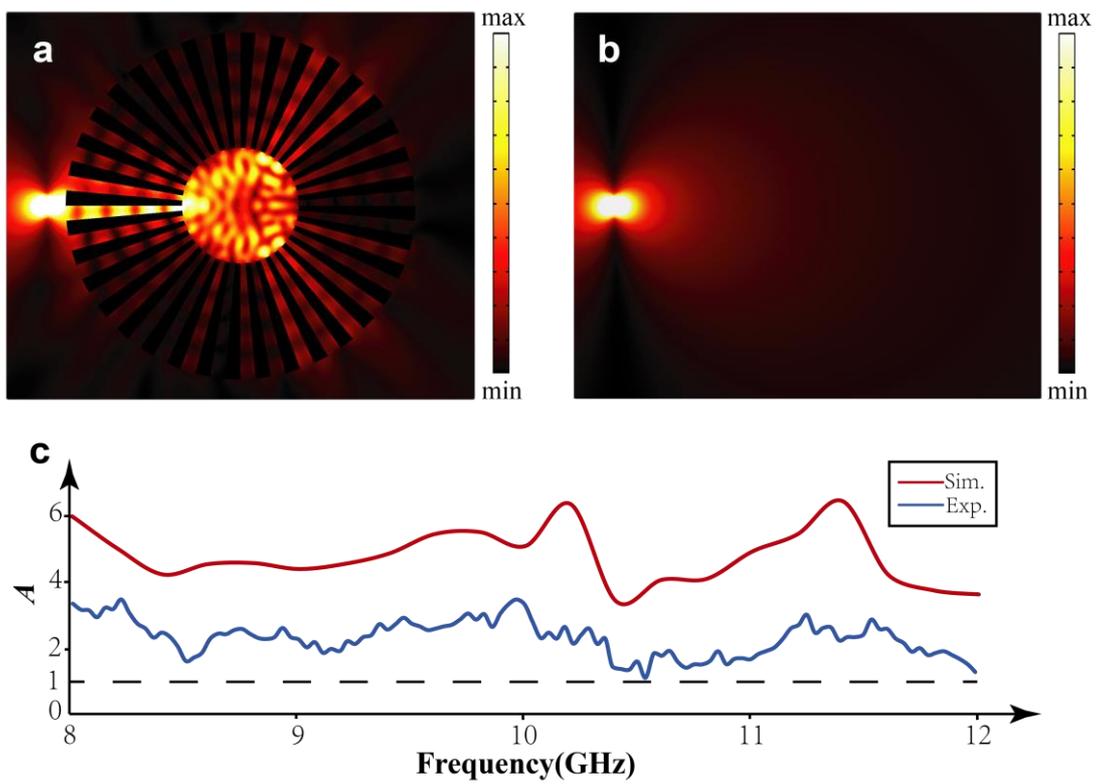